\documentclass[a4paper,fleqn]{cas-dc}

\usepackage[numbers]{natbib}
\usepackage{amsmath}
\usepackage{graphicx}
\usepackage{braket}
\usepackage{bm} 
\usepackage{listings}
\usepackage{amssymb}
\usepackage{subfig}
\usepackage{dcolumn}
\usepackage{xcolor}
\def\tsc#1{\csdef{#1}{\textsc{\lowercase{#1}}\xspace}}
\tsc{WGM}
\tsc{QE}
\tsc{EP}
\tsc{PMS}
\tsc{BEC}
\tsc{DE}

\begin{document}
\let\WriteBookmarks\relax
\def\floatpagepagefraction{1}
\def\textpagefraction{.001}

\shorttitle{Transitionless phonon assisted photon-qubit quantum state transfer in a hybrid optomechanical system}

\shortauthors{A. Saha et~al.}

\title[mode = title]{Transitionless phonon assisted photon-qubit quantum state transfer in a hybrid optomechanical system}

\author[1]{Arindam Saha}[
    orcid=0000-0002-3120-3375]
  \ead{arindam96@outlook.com} 
  
\author[1] {Amarendra K. Sarma}[
    orcid=0000-0003-4846-0893]  
\ead{aksarma@iitg.ac.in}

\address[1]{Department of Physics, Indian Institute of Technology Guwahati, Guwahati-781039, Assam, India}

\begin{abstract}
Quantum state transfer is crucial for quantum information processing and quantum computation. Here, we propose a hybrid optomechanical system capable of coupling a qubit, an optical mode and a mechanical oscillator. The displacement of the mechanical oscillator,due to radiation pressure, induces spatial variation in the cavity field, which in turn couples the qubit with phonon mode. This allows state transfer between the cavity mode and the qubit without an actual interaction between them. We present a scheme for the transitionless quantum state transfer based on the transitionless quantum driving algorithm allowing us to achieve perfect state transfer.
\end{abstract}

\begin{keywords}
Cavity QED \sep Optomechanics \sep Light-matter Interaction
\end{keywords}

\maketitle

\section{Introduction}
High fidelity quantum state transfer (QST) is essential for the implementation of scalable quantum computation and quantum communication protocols \cite{intro.a}. This is the central goal in various schemes related to quantum networks \cite{intro.b,r1}. Numerous quantum state transfer schemes and protocols have been proposed and studies in many physical systems \cite{intro.b,s1,s2,s3,s4}. Recently,  quantum optomechanical systems (OMS) have drawn tremendous research interest due to its potential as a promising platform for quantum information processing related applications \cite{r1,intro.c}. Quantum networks require interfacing stationary qubits that store and process information and photons for communication. Optomechanics facilitates such situation by providing a free choice of photon wavelength, allowing us to use telecom band for long-distance communication \cite{new2}. In particular, hybrid quantum optomechanical systems owing to the versatility of optical and mechanical components in coupling to different systems such as spins, cold atoms, superconducting qubits, etc. are highly in focus \cite{r1,r2,r3,r4,new1,new4}. It's quite clear that quantum state transfer between various modes in OMS is a topic of great research interest \cite{r5,r6}. QST is usually achieved by adiabatic evolution of the dark eigenstate, done mostly by the so-called stimulated Raman adiabatic passage (STIRAP) protocols \cite{r7}. However the process requires considerable operational time to satisfy the adiabatic condition, which leads to decoherence \cite{intro.c}. Recently various shortcut to adiabaticity (STA) approaches have been proposed, which allows one to cancel the unwanted transitions between eigenstates by applying precisely controlled external fields \cite{intro.c}. One such approach is the so-called transitionless quantum driving (TQD). The key idea is to add a counter-diabatic term to the original Hamiltonian to get the one that drives the states transitionlessly. The TQD process involves simple calculation and is easy to implement in experiments. It has been experimentally demonstrated to produce high fidelity state transfer, with low dependence on controlling parameters and robust to mechanical dissipation \cite{tqd1,tqd2}. In this work, we propose a phonon mediated quantum state transfer scheme to have a fast and robust state transfer between a photon and a qubit.

\begin{figure}
	\begin{center}
		\includegraphics[width=0.47\textwidth]{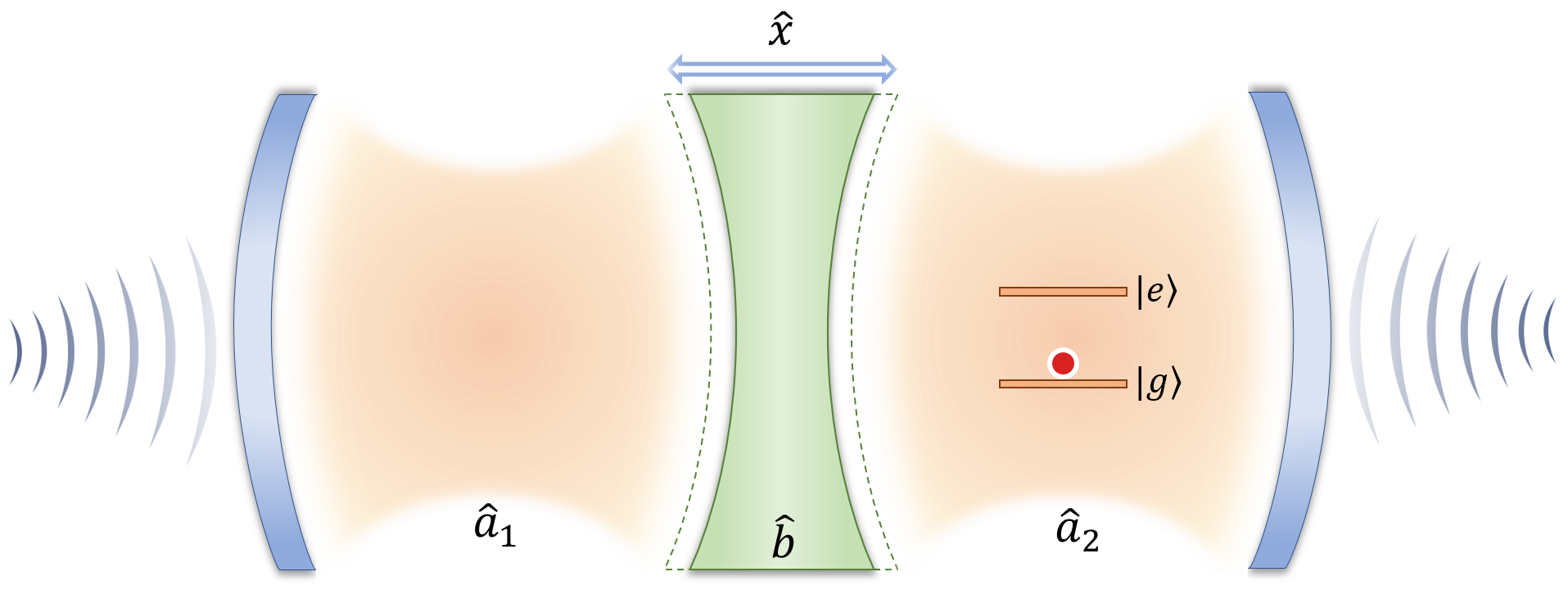}
	\end{center}
	\caption{Schematic of the hybrid optomechanical system. The components are represented as, electric field (maroon), oscillator (green), atom (red)}
	\label{fig1}
\end{figure}

\section{The model}
We consider a cavity QED set-up with a membrane/ mirror in between, as depicted in Fig. \ref{fig1}. It consists of two optical cavities with frequencies $\omega_{c1}$ and $\omega_{c2}$, $\hat{a}_1$ and $\hat{a}_2$ being their respective annihilation operators. The optical cavities are connected to each other by a oscillating mirror/ membrane of frequency $\omega_m$ in between them, $\hat{b}$ being the annihilation operator of the mechanical/phonon mode. The second cavity consists of a two level atom/ qubit with frequency $\omega_A$ in it. The Hamiltonian of the system \cite{sys.a,sys.b} could be written as $(\hbar=1)$,
\begin{equation}
\begin{split}
    H = \omega_{c1}\hat{a}_1^{\dagger}\hat{a}_1 +\omega_{c2} \hat{a}_2^{\dagger}\hat{a}_2 + \frac{\omega_A}{2}\hat{\sigma}_z + \omega_m \hat{b}^{\dagger}\hat{b} - g_1'(\hat{b}+\hat{b}^{\dagger})\hat{a}_1^{\dagger}\hat{a}_1\\ - g_2'(\hat{b}+\hat{b}^{\dagger})\hat{a}_2^{\dagger}\hat{a}_2 + g'(\hat{a}_2+\hat{a}_2^{\dagger})(\hat{\sigma}_+ + \hat{\sigma}_-)
    \label{eq.1}
\end{split}
\end{equation}

where $\hat{\sigma}_+=\left|e \right\rangle \left\langle g\right| $, $\hat{\sigma}_-=\left|g \right\rangle \left\langle e\right|$ and $\hat{\sigma}_z=\left|e \right\rangle \left\langle e\right| - \left|g \right\rangle \left\langle g\right|$ are the operators describing the qubit. The single-photon optomechanical coupling strengths between the mode $\hat{a}_1$ $(\hat{a}_2)$ and $\hat{b}$ is described by $g_1'$ $(g_2')$, while the atom-photon coupling is denoted by $g'$. For convenience, we switch to the frame rotating with laser frequency $\omega_L$. By applying unitary transformation, $\hat{U}=exp(-i\omega_L(\hat{a}_1^{\dagger}\hat{a}_1+\hat{a}_2^{\dagger}\hat{a}_2)t)$, and generating the new Hamiltonian as, $H=\hat{U}H_{old}\hat{U}^{\dagger}-i\hat{U}\partial\hat{U}^{\dagger}/\partial t$, we have,
\begin{equation}
\begin{split}
    H = \Delta_{c1}\hat{a}_1^{\dagger}\hat{a}_1 +\Delta_{c2} \hat{a}_2^{\dagger}\hat{a}_2 + \frac{\omega_A}{2}\hat{\sigma}_z + \omega_m \hat{b}^{\dagger}\hat{b} - g_1'(\hat{b}+\hat{b}^{\dagger})\hat{a}_1^{\dagger}\hat{a}_1\\ - g_2'(\hat{b}+\hat{b}^{\dagger})\hat{a}_2^{\dagger}\hat{a}_2 + g'(\hat{a}_2+\hat{a}_2^{\dagger})(\hat{\sigma}_+ + \hat{\sigma}_-)
    \label{eq.2}
\end{split}
\end{equation}

where, $\Delta_{ci}=\omega_{ci}-\omega_L$, with $i={1,2}$. In the interaction picture, the Hamiltonian in (\ref{eq.2}), is transformed by applying unitary transformation $\hat{U}=exp(-iH_0t)$ with $H_0=\Delta_{c1}(\hat{a}_1^{\dagger}\hat{a}_1 + \hat{a}_2^{\dagger}\hat{a}_2 + \hat{\sigma}_z/2)$. Hence, we obtain,
\begin{equation}
\begin{split}
    H = \Delta_c \hat{a}_2^{\dagger}\hat{a}_2+\frac{\omega_A-\Delta_{c1}}{2}\hat{\sigma}_z + \omega_m \hat{b}^{\dagger}\hat{b} - g_1'(\hat{b}+\hat{b}^{\dagger})\hat{a}_1^{\dagger}\hat{a}_1\\ - g_2'(\hat{b}+\hat{b}^{\dagger})\hat{a}_2^{\dagger}\hat{a}_2 + g'(\hat{a}_2+\hat{a}_2^{\dagger})(\hat{\sigma}_+ + \hat{\sigma}_-)
    \label{eq.3}
\end{split}
\end{equation}

where $\Delta_c = \Delta_{c2}-\Delta_{c1}$. Now we apply mean field approximation to the cavity modes as $\hat{a}_i=\sqrt{n_i}+\delta\hat{a}_i$, with $i={1,2}$. We consider pumping the second cavity with a strong field such that $n_2 \gg 1$ and we can neglect its fluctuations as $\delta \hat{a}_2 \to 0$. Next, we consider the situation where the displacement of the central membrane causes variation in the cavity field distribution. As a result, the atom-cavity coupling rate becomes dependent on the position of the oscillator \cite{sys.a, sys.b}. Thus we may write the coupling rate as $g'(\hat{x})=g'(0)+(\partial g'/\partial x)|_{x=0} \hat{x}$, where $\hat{x}=x_{ZPF}(b+b^{\dagger})$. In case of mechanical equilibrium $g'(0)=0$. Therefore, eliminating the non-resonant interaction terms under RWA, the Hamiltonian in (\ref{eq.3}) could be rewritten as: 

\begin{equation}
\begin{split}
    H=\Delta_c n_2+\frac{\omega_A-\Delta_{c1}}{2}\hat{\sigma}_z + \omega_m \hat{b}^{\dagger}\hat{b} + g_1(\delta \hat{a}_1^{\dagger}\hat{b}+\delta\hat{a}_1\hat{b}^{\dagger})\\ - g_2'(\hat{b}+\hat{b}^{\dagger})n_2 + g(\hat{b}\hat{\sigma}_+ + \hat{b}^{\dagger}\hat{\sigma}_-)
    \label{eq.4}
\end{split}
\end{equation}

where $g_1=g_1'\sqrt{n_1}$ and $g=2\sqrt{n_2}(\partial g'/\partial x)|_{x=0}x_{ZPF}$. The term $- g_2'(\hat{b}+\hat{b}^{\dagger})n_2$ denotes average radiation pressure field and can be neglected by shifting the displacement's origin and modifying the detuning $\Delta_c$ by adding a constant term to it \cite{r1}. The term $\Delta_c n_2$ along with the term added to it are constants and can be safely neglected without affecting the dynamics of the system. Further, the cavity-qubit and the optomechanical coupling parameters can be related as follows. We have $g_1=\sqrt{n_1}(\partial \omega_{c1}/\partial x)x_{ZPF} $ for the first cavity, while $g_2=\sqrt{n_2}(\partial \omega_{c2}/\partial x)x_{ZPF} $ for the second cavity. Again, we know $g'=\alpha \sqrt{\omega_{c2}}$, $\alpha$ being some constant \cite{b1}. Therefore, $g=(\alpha/\sqrt{\omega_{c2}})\sqrt{n_2} (\partial \omega_{c2}/\partial x)x_{ZPF}=\gamma g_2$, $\gamma$ being some other constant, which signifies the strength of the coupling. Thus eliminating the $\hat{a}_2$ mode completely from the equation, and writing: $\delta\hat{a}_1\equiv \hat{a}$ and $\Delta_{c1}\equiv \delta_c$, the Hamiltonian in (\ref{eq.4}) could finally be put in the following form:
\begin{equation}
\begin{split}
    H=\frac{\omega_A-\delta_c}{2}\hat{\sigma}_z + \omega_m \hat{b}^{\dagger}\hat{b} + g_1(\hat{a}^{\dagger}\hat{b}+\hat{a}\hat{b}^{\dagger})\\ + \gamma g_2(\hat{b}\hat{\sigma}_+ + \hat{b}^{\dagger}\hat{\sigma}_-)
    \label{eq.5}
\end{split}
\end{equation}

In this work, we consider the following Vitanov-type time-dependent coupling \cite{sys.c} parameters:
$$g_1(t)=\sin(\theta(t))$$
$$g_2(t)=\cos(\theta(t))$$ $$\theta(t)=\frac{\pi}{2}\frac{1}{1+e^{-v(t-3/v)}}$$

Here, $v$ controls rapidity of the coupling strength. It is worthwhile to note that the Vitanov-type of coupling is used to optimize the adiabatic passage like, STIRAP, to maximise fidelity and minimise non-adiabatic transitions \cite{r8}.

\section{Results and Discussions}
Adiabatic passage method is the most used tool in atom-optics and condensed matter physics to obtain efficient population transfer as well as in state transfer related protocols. However in the set-up considered in this work, this method fails to yield useful results. In order to investigate state transfer under adiabatic conditions, we evolve the system with the Hamiltonian (\ref{eq.5}) taking $\omega_m = \omega_A-\delta_c = 1$ GHz and $\gamma=20$ for different values of $v$.
\begin{figure}
		\centering
		\subfloat{{\includegraphics[width=0.245\textwidth]{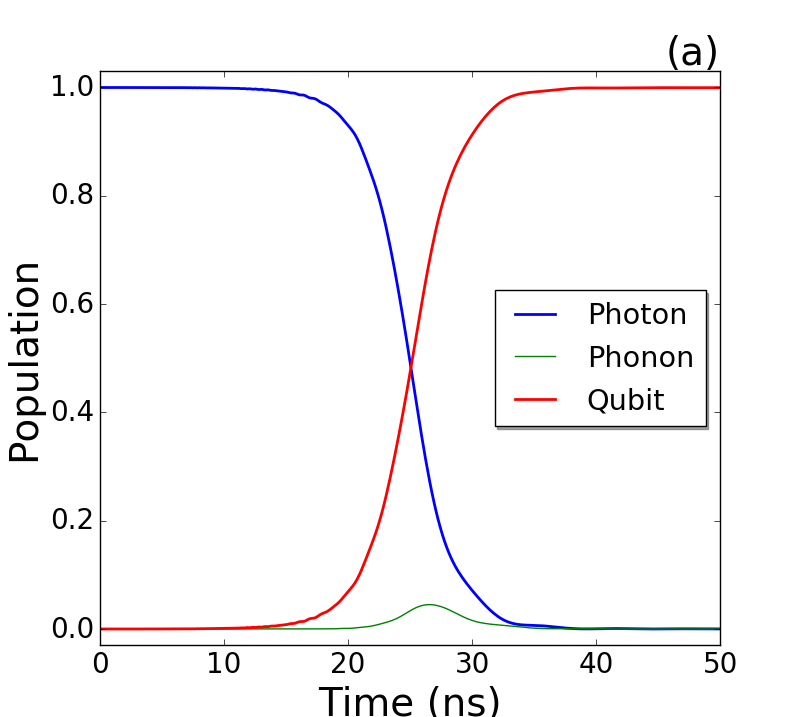}}}%
		\subfloat{{\includegraphics[width=0.245\textwidth]{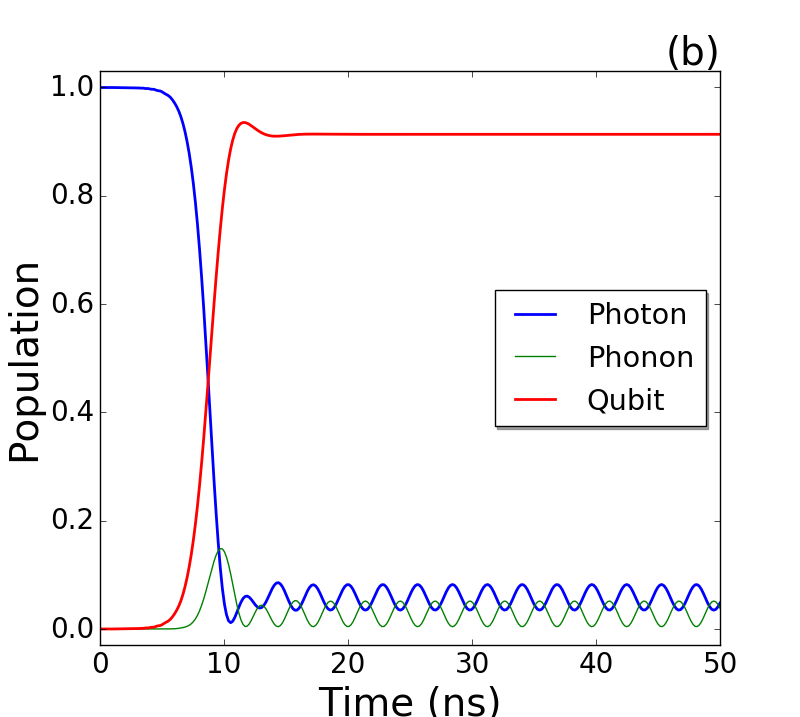}}}\\
		\subfloat{{\includegraphics[width=0.245\textwidth]{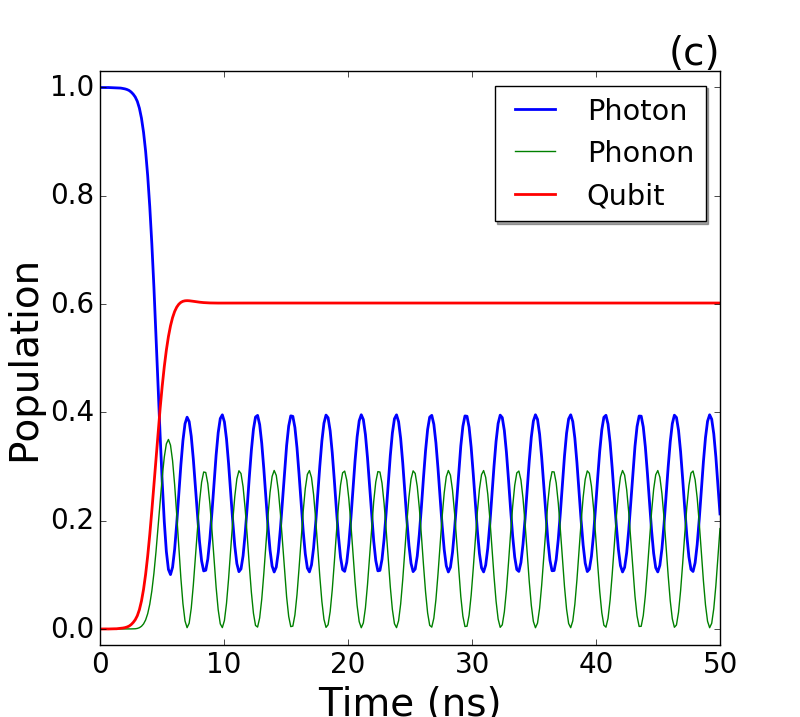}}}%
	    \subfloat{{\includegraphics[width=0.245\textwidth]{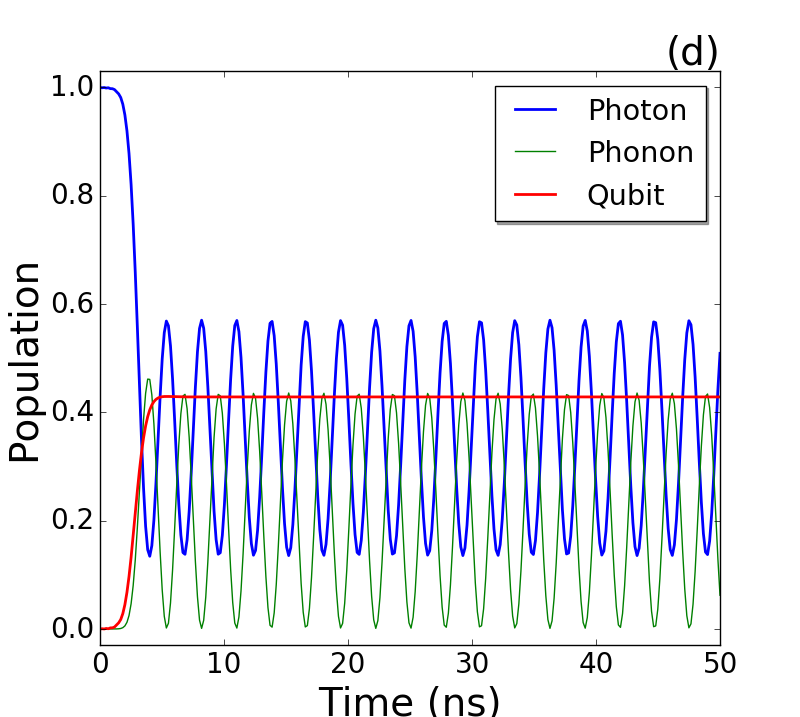}}}%
	   \caption{Simulation of quantum state transfer for the following values of $v$, (a) 0.25, (b) 0.75, (c) 1.5, (d) 2 }
	   \label{fig.2}
\end{figure}

The results are depicted in Fig. \ref{fig.2}. It could be observed that, when the coupling strength is varied slowly/adiabatically as in case of $v=0.25$, we can achieve an almost perfect population transfer. However when it evolved rapidly/non-adiabatically, the system transits into intermediate states and fails to achieve a perfect population transfer. In order to solve this problem, we apply the transitionless quantum driving (TQD) algorithm \cite{text.b} to the system. In the following, we briefly discuss the TQD algorithm. 

\subsection{ Transitionless Quantum Driving algorithm}
 	When a Hamiltonian $\hat{H}_0(t)$ is time dependent, it induces transition between the quantum states. If the change is slow, the adiabatic theorem states that the system will follow the instantaneous eigenstates it started with. But there still remains a very small but finite transition probability. However it is possible to find a $\hat{H}(t)$ associated with $\hat{H}_0(t)$ that can drive the eigenstates of $\hat{H}_0(t)$ exactly without transition between them, even in the non adiabatic regime \cite{text.b, gtqd}. Lets consider an time-dependent Hamiltonian  $\hat{H}_0(t)$, that drives its instantaneous eigenstates $\ket{n(t)}$ with energies $E_n(t)$ as,
 	\begin{equation}
 	    \hat{H}_0(t)\ket{n(t)} = E_n(t)\ket{n(t)} 
 	\end{equation}
 	
 	 It follows from the adiabatic theorem that the states driven by $H_0(t)$ can be expressed as,
 	 \begin{multline}
 	    \ket{\varphi_n(t)}=exp \biggl( -i\int_{0}^{t}dt'E_n(t') -\\ \int_{0}^{t}dt' \braket{n(t')|\dot{n}(t')} \biggr) \ket{n(t)}
 	 \end{multline}
 	 
 	The new Hamiltonian $\hat{H}(t)$ that drives these evolving states exactly without transition among them should satisfy,
 	\begin{gather}
 	    i\partial_t\ket{\varphi_n(t)} =\hat{H}(t)\ket{\varphi_n}
 	\end{gather}
 	
 	Moreover any time-dependent operator $\hat{U}(t)$ will also satisfy the above equation
 	\begin{equation}
 	    i\partial_t\hat{U}(t) =\hat{H}(t)\hat{U}(t)
 	\end{equation}
 	
 where $\hat{H}(t)=i\left(\partial_t\hat{U}(t)\right)\hat{U}^{\dagger}(t)$. In order to ensure that no transition takes place we choose,
 \begin{multline}
    \hat{U}(t)=\sum_{n}exp \biggl( -i\int_{0}^{t}dt'E_n(t') -\\ \int_{0}^{t}dt' \braket{n(t')|\dot{n}(t')} \biggr) \ket{n(t)} \bra{n(0)}  
 \end{multline}
 
Using this we find our required Hamiltonian driving eigenstates $\ket{n(t)}$ as,
\begin{equation}
   \hat{H}(t)=\sum_n E_n \ket{n} \bra{n} + i\sum_n\big(\ket{\dot{n}} \bra{n} - \braket{n|\dot{n}} \ket{n} \bra{n}\big)  
\end{equation}

As a result we obtain the most general form of Hamiltonian that drives the eigenstates without causing transition between them. The first term represents the interaction Hamiltonian that drives the system in the adiabatic regime, and the second term is the counter-diabatic Hamiltonian that denotes a shortcut to the adiabaticity.

\subsection{Population transfer via the TQD algorithm}
To begin with, we take the interacting part of the Hamiltonian (\ref{eq.4}) as follows:
\begin{equation}
    H_I= g_1(\hat{a}^{\dagger}\hat{b}+\hat{a}\hat{b}^{\dagger})+ \gamma g_2(\hat{b}\hat{\sigma}_+ + \hat{b}^{\dagger}\hat{\sigma}_-)
    \label{eq.6}
\end{equation}

We consider the following basis vectors, represented as\\ $\left(\ket{\phi}=\ket{\hat{a}} \ket{\hat{b}} \ket{\hat{\sigma}}\right)$:
$$\ket{\phi_1}=\ket{1} \ket{0} \ket{g}$$
$$\ket{\phi_2}=\ket{0} \ket{1} \ket{g}$$
$$\ket{\phi_3}=\ket{0} \ket{0} \ket{e}$$

The Hamiltonian is then expanded in the above basis, and in the matrix form, the Hamiltonian (\ref{eq.6}) takes the following form:
\begin{equation}
	H_{I}(t) = 
	\begin{bmatrix}
    0 & g_1(t) & 0\\
	g_1(t) & 0 & \gamma g_2(t)\\
	0 & \gamma g_2(t) & 0
	\end{bmatrix} 
	\label{eq.7}
\end{equation}
It has the following eigenfunctions,
    $$\ket{\psi_1}=-\frac{\gamma g_2}{g_0}\ket{\phi_1}+\frac{g_1}{g_0}\ket{\phi_3}$$
    $$\ket{\psi_2}=\frac{1}{\sqrt 2}\left[\frac{g_1}{g_0}\ket{\phi_1}-\ket{\phi_2}+\frac{\gamma g_2}{g_0}\ket{\phi_3}\right]$$
    $$\ket{\psi_3}=\frac{1}{\sqrt 2}\left[\frac{g_1}{g_0}\ket{\phi_1}+\ket{\phi_2}+\frac{\gamma g_2}{g_0}\ket{\phi_3}\right]$$
    
with eigenvalues $0,-g_0$ and $g_0$ respectively, where $g_0=\sqrt{g_1^2+\gamma^2 g_2^2}$. On applying the TQD algorithm \cite{text.b, gtqd}, we derive the following counter-diabatic Hamiltonian,
\begin{equation}
\begin{split}
    H_{TQD} & =i\sum_{n=1}^3\left(\ket{\dot{\psi}_n}\bra{\psi_n} - \braket{\psi_n|\dot{\psi}_n} \ket{\psi_n} \bra{\psi_n}\right) \\
    & = iG\left(\ket{\phi_1}\bra{\phi_3}- \ket{\phi_3}\bra{\phi_1}\right)
 	\label{eq.8}
\end{split}
\end{equation}

with $G=\gamma(\dot{g_1}g_2-g_1\dot{g_2})/g_0^2$. Hence, using the above result we get the new Hamiltonian that drives the states without transition as,
\begin{equation}
    \mathcal{H}=H_{I}+H_{TQD}
\end{equation}

\begin{figure}
		\centering
		\subfloat{{\includegraphics[width=0.245\textwidth]{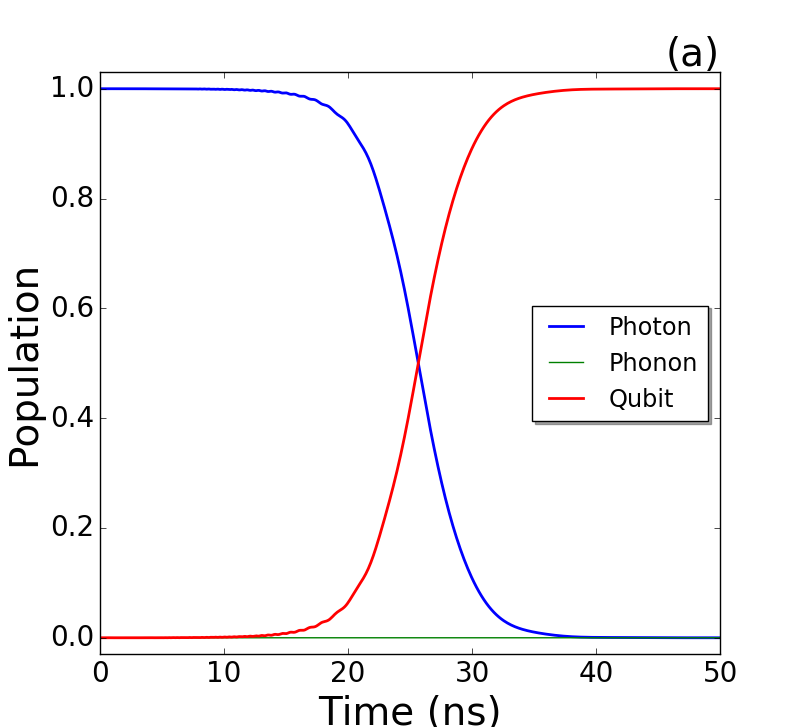}}}%
		\subfloat{{\includegraphics[width=0.245\textwidth]{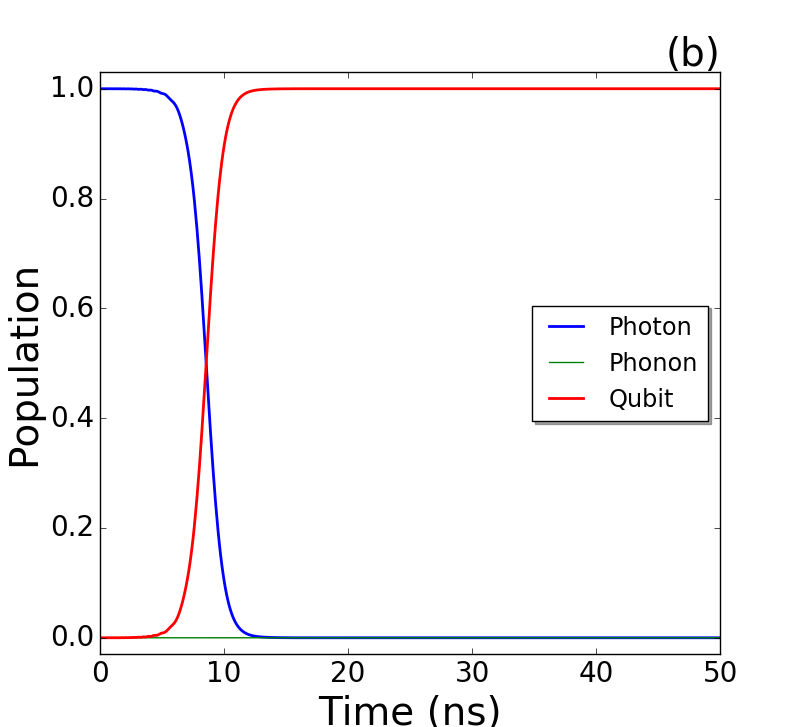}}}\\
		\subfloat{{\includegraphics[width=0.245\textwidth]{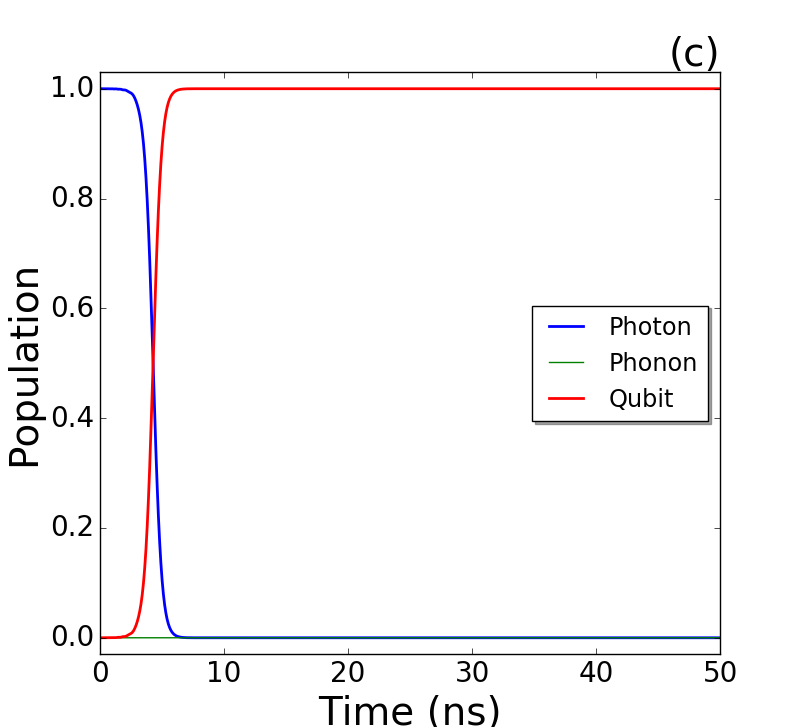}}}%
	    \subfloat{{\includegraphics[width=0.245\textwidth]{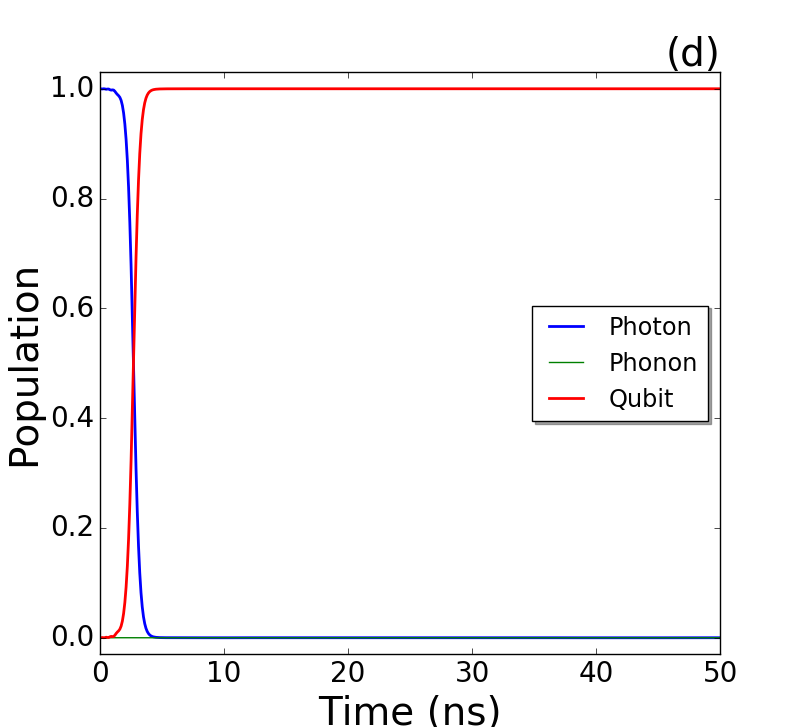}}}%
	   \caption{Simulation of quantum state transfer with TQD algorithm for the following values of v, (a) 0.25, (b) 0.75, (c) 1.5, (d) 2 }
	   \label{fig.4}
\end{figure}

Fig. \ref{fig.4} depicts the simulation of quantum state transfer with this new Hamiltonian, i.e. the TQD algorithm for various values of the $v$-parameter. It is worth perceiving that the counter-diabatic Hamiltonian evolves the initial state $\ket{\phi_1}$, associated with the dark eigenstate $\ket{\psi_1}$. Further it restricts transition to other eigenstates even in the presence of all forms of interaction. As a consequence, this confines the occupation to the states $\ket{\phi_1}$ and $\ket{\phi_3}$, keeping the optomechanical mode completely unperturbed. The population transfer thus happens by controlling the coupling parameters, $g_1$ and $g_2$, precisely. Hence, it is evident that, one can now seamlessly transfer population from $\ket{\phi_1}$ to $\ket{\phi_3}$ state without exciting $\ket{\phi_2}$ state in the process, making it transitionless. We find that, even when the duration of the coupling strength is reduced, it is possible to obtain very fast population transfer with no attenuation. Hence, the parameter $v$ could be used to control the rapidity of the population transfer.

\subsection{Fidelity of state transfer}
\begin{figure}
		\centering
		\subfloat{{\includegraphics[width=0.245\textwidth]{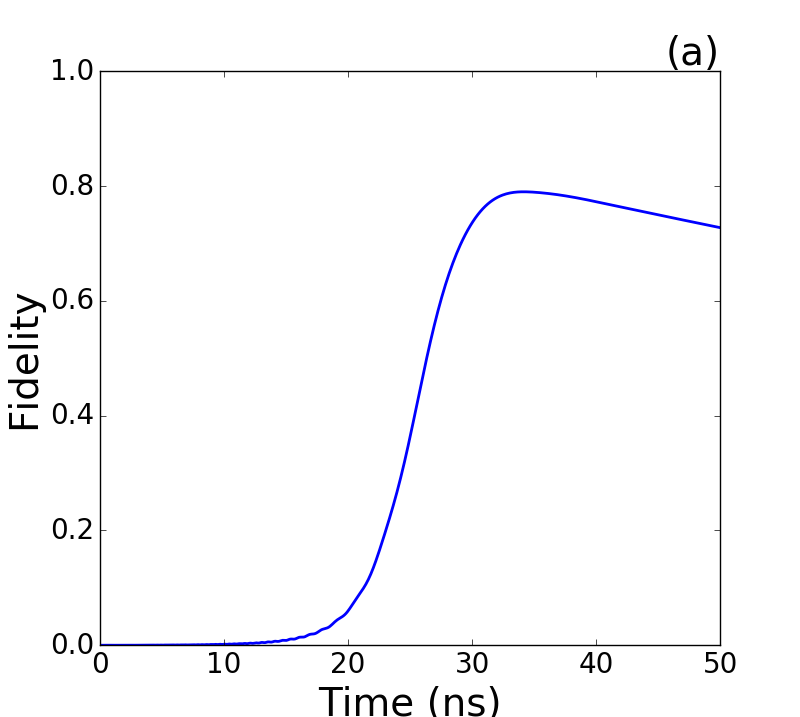}}}%
		\subfloat{{\includegraphics[width=0.245\textwidth]{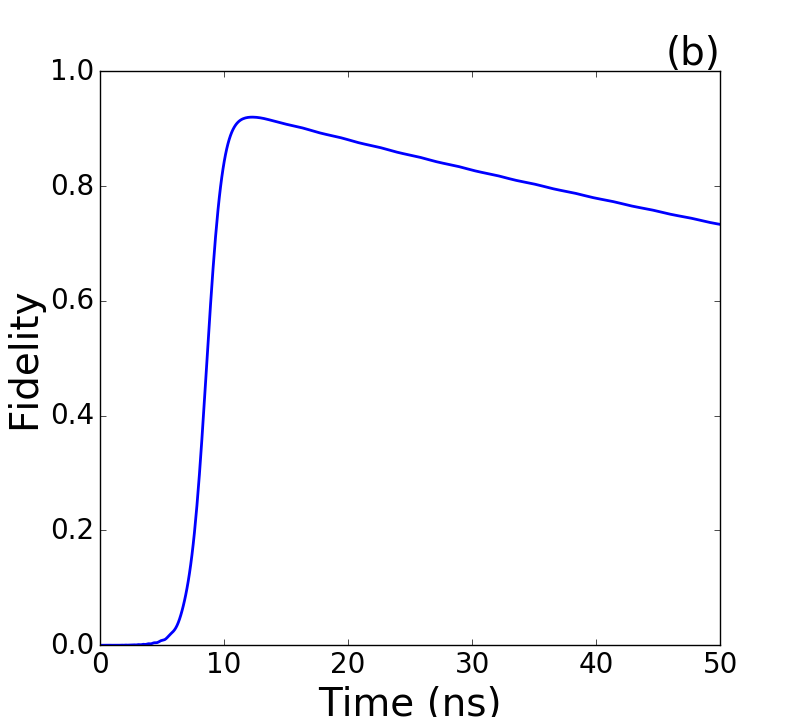}}}
		\caption{Fidelity vs time  (a) $v=0.25$ (b) $v=0.75$. $\kappa=0.005 GHz$, $\Gamma = 0.005 GHz$, $\gamma_m = 0.05 MHz$ and $n_{th} = 50$ }
		\label{fig.5}
\end{figure}

\begin{figure}
	\begin{center}
		\includegraphics[width=0.42\textwidth]{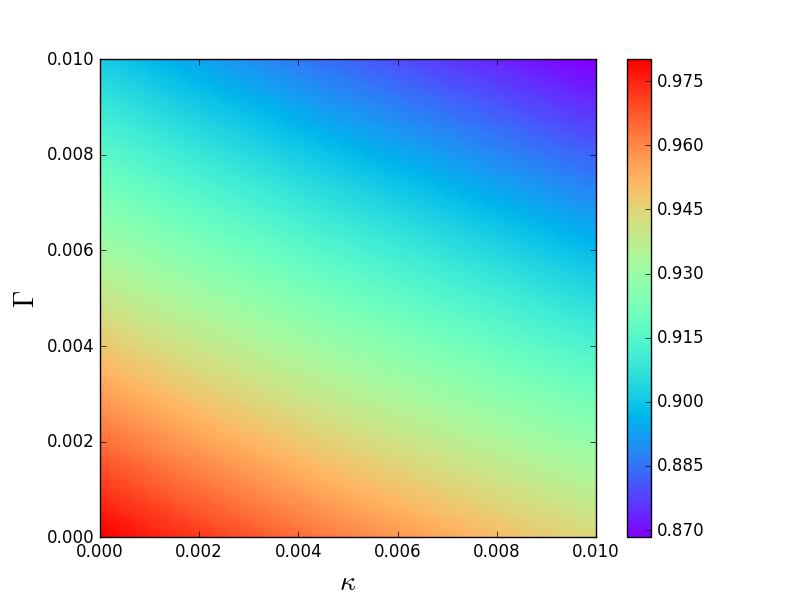}
	\end{center}
	\caption{Dependence of fidelity on cavity $\kappa$ and qubit $\Gamma$ decay rates (in $GHz$) for $v = 0.75$ and $\gamma_m = 0.05 MHz$}
	\label{fig.6}
\end{figure}

The process discussed till now was lossless. When we consider a non conservative interaction, one needs to take into account the finite decay rates of the cavity, mechanical mode and the qubit. This could be done by considering the so-called quantum master equation. To study the dynamics, we solve the master equation \cite{text.c} of the following form,
\begin{multline}
    \frac{\partial \rho}{\partial t}=-i[\mathcal{H},\rho]+\kappa L[\hat{a}]\rho + \Gamma L[\hat{\sigma}_-] \rho \\+ \gamma_m (n_{th}+1) L[\hat{b}] \rho + n_{th} \gamma_m L[\hat{b}^{\dagger}]\rho
\end{multline}

where $L[A]\rho=\left[ 2A\rho A^{\dagger} - A^{\dagger}A\rho - \rho A^{\dagger}A\right]/2$. $\kappa$ and $\gamma_m$ are the decay rates of the cavity and the mechanical oscillator respectively, $\Gamma$ is the decay rate of qubit and $n_{th}$ the average number of phonons in the external bath. Further, the fidelity of state transfer is calculated as $F=\bra{01}tr_m(\rho)\ket{01}$, where $tr_m(\rho)$ is the reduced density matrix, after tracing out the mechanical mode. $\ket{01}$ denotes the state where there is 0 photon in cavity 1, while the qubit is in the excited state. The results are illustrated in Fig. \ref{fig.5}. Also the dependence of the fidelity with the cavity and atom decay rates is shown in Fig. \ref{fig.6}.

It is quite clear that the proposed quantum state transfer protocol is reasonably robust even in the presence of cavity and atom decay achieving a fidelity of more than $90\%$. Hence, it is possible to seamlessly transfer a quantum state from photon to a qubit mediated by phonons with the proposed method. Moreover, it has been demonstrated that these form of interaction is further assisted by the high phonon-qubit coupling strength than the usual photon-qubit one \cite{new3}. Experimentally, the proposed model may be achieved by using independently controllable coupling parameters. The optomechanical coupling strength is modulated by varying the length of the cavity. For instance, this can be achieved, as suggested in \cite{exp.a} by varying the dimpled fibre taper attached to the central mirror by some nanopositioner. The lengths of the two cavities could be independently modulated to result in different $g_1(t)$ and $g_2(t)$. Moreover, to keep the photon number high in the second cavity, it can be irradiated with an intense laser. To execute the TQD process, an additional field is applied, and the coupling parameters are varied according to $G(t)$ to simulate $H_{TQD}$ \cite{exp.b}.

 \section{Conclusions}
We have proposed a method to transfer quantum state from photon to a qubit mediated by phonons with high fidelity. The displacement of the mechanical oscillator, due to radiation pressure, induces spatial variation in the cavity field, which in turn couples the qubit with phonon mode. This allows state transfer between the cavity mode and the qubit without an actual interaction between them. A population transfer through the usual adiabatic process either takes longer duration or suffers from population leakage, both of which ultimately amounts to a significant loss. The scheme based on the transitionless quantum driving algorithm makes the process faster and robust against decoherence. It further increases its fidelity with the rate of coupling. The discussed model confers some unique advantages useful for implementation in quantum communication and computation, and TQD assists in fine-tuning them.

\bibliographystyle{model1-num-names}
\bibliography{mybibfile}

\end{document}